
\input harvmac

\def\np#1#2#3{Nucl. Phys. {\bf B#1} (#2) #3}
\def\pl#1#2#3{Phys. Lett. {\bf #1B} (#2) #3}

\def\physrev#1#2#3{Phys. Rev. {\bf D#1} (#2) #3}

\def\prep#1#2#3{Phys. Rep. {\bf #1} (#2) #3}

\def\cmp#1#2#3{Comm. Math. Phys. {\bf #1} (#2) #3}

\def\Tr{{\rm Tr ~}}

\def\cn{{\cal N}}

\def\tilde{\widetilde}

\Title{hep-th/9505004, EFI-95-20 , WIS/4/95 }
{\vbox{\centerline{On Duality in Supersymmetric Yang-Mills Theory}
}}
\bigskip
\centerline{\it D. Kutasov}
\smallskip
\centerline{Enrico Fermi Institute and}
\centerline{Department of Physics}
\centerline{University of Chicago}
\centerline{Chicago, IL 60637, USA}

\vskip .2in
\centerline{\it A. Schwimmer}
\smallskip\centerline
{Department of Physics of Complex Systems}
\centerline{Weizmann Institute of Science}
\centerline{Rehovot, 76100, Israel}
\vskip .2in

\vglue .3cm

\bigskip\bigskip

\noindent
We discuss non-abelian $SU(N_c)$ gauge theory coupled to
an adjoint chiral superfield $X$, and a number of fundamental
chiral superfields $Q^i$. Using duality, we show that turning
on a superpotential $W(X)=\Tr\sum_{l=1}^k g_l X^{l+1}$ leads to
non-trivial long distance dynamics, a large number of
multicritical IR fixed points and vacua, connected
to each other by varying the coefficients $g_l$.
\Date{4/95}

\nref\ads{I. Affleck, M. Dine, and N. Seiberg, \np{241}{1984}{493};
\np{256}{1985}{557}}%
\nref\nsvz{V.A. Novikov, M.A. Shifman, A. I.  Vainstein and V. I.
Zakharov, \np{223}{1983}{445}; \np{260}{1985}{157};
 \np{229}{1983}{381}}
\nref\svholo{M.A. Shifman and A.I Vainshtein, \np{277}{1986}{456};
\np{359}{1991}{571}}%
\nref\cern{D. Amati, K. Konishi, Y. Meurice, G.C. Rossi and G.
Veneziano, \prep{162}{1988}{169} and references therein}%
\nref\nonren{N. Seiberg,                 \pl{318}{1993}{469}}%
\nref\natii{N. Seiberg,                 \physrev{49}{1994}{6857}}%
\nref\ils{K. Intriligator, R. Leigh and N. Seiberg,
\physrev{50}{1994}{1092}
}%
\nref\sw{N. Seiberg and E. Witten, \np{426}{1994}{19};
                \np{431} {1994} {484}
}%
\nref\intse{K. Intriligator and N. Seiberg,
\np{431} {1994} {551}
}%
\nref\iss{K. Intriligator, N. Seiberg and S. Shenker,
\pl{342}{1995}{152}}%
\nref\nati{N. Seiberg, \np {435}{1995}{129}}%
\nref\ni{K. Intriligator and N. Seiberg, RU-95-3, IASSNS-HEP-95/5,
hep-th/9503179}%
\nref\dk{D. Kutasov, EFI-95-11, hep-th/9503086}%
\newsec{Introduction.}

The recent progress in understanding the role of holomorphicity
\refs{\ads -\iss}\ and duality \refs{\nati-\dk}\
in four dimensional supersymmetric field theories can be used in many cases
to study strongly coupled theories. In this note we discuss
a class of theories where strong coupling effects lead to a rich
pattern of fixed points exhibiting new duality symmetries.

We study supersymmetric Yang-Mills theory with gauge group $SU(N_c)$,
a chiral matter superfield $X$ in the adjoint representation
of the gauge group and $N_f$ fundamental multiplets $Q^i$ accompanied
by $N_f$ anti--fundamental multiplets $\tilde Q_i$, $i=1,\cdots, N_f$.
This theory is asymptotically free for $N_f<2N_c$.
The model without a superpotential for the matter fields is governed
by a non-trivial infrared fixed point, and is in a non abelian Coulomb
phase
(for $N_f>0$). This interesting model has so far resisted
all attempts at a detailed understanding. We will instead discuss the
model with a superpotential\foot{We consider $k<N_c$
for simplicity.},
\eqn\b{W=g_k\Tr X^{k+1}.}
Adding the superpotential \b\ has the following consequences:

\noindent a) If the operator $\Tr X^{k+1}$ is relevant at the infrared
fixed point of the theory with $W=0$, adding \b\ drives the system to a
new fixed point. For $k>2$, $\Tr X^{k+1}$ is irrelevant near the UV fixed
point.
We will argue below that there is a range (which depends on $k$)
of $N_f$ for which $\Tr X^{k+1}$
is relevant in the IR. For given $N_c>k>2$ there is a critical
number of flavors $N_0(k, N_c)$,
$2N_c/k<N_0(k,N_c)<2N_c$ such that if
$N_f<N_0(k,N_c)$ the operator $X^{k+1}$ is relevant in the IR limit of
the $W=0$ theory. Conversely, for given $N_f<2N_c$ all operators $\Tr
X^{k+1}$
with $k\le k_0(N_f,N_c)$ are relevant $(k_0>2)$ in the infrared.

\noindent b) Whether or not adding \b\ leads to a new fixed point, one
can not ignore this superpotential. It has the effect of lifting the flat
directions of the theory with no superpotential corresponding to giving
$X$ an expectation value. In addition, the superpotential \b\ leads to
a truncation of the chiral ring.
The equations of motion for
$X$ set:
\eqn\h{X^k-{1\over N}(\Tr X^k){\bf 1}={\rm D\;\; term}}
Hence the chiral operators involving $X$ in the presence of the
superpotential \b\ are $\Tr X^l$, $l=2,\cdots, k$,
and operators involving the matrix $X^l$ with $l<k$. There are two kinds
of gauge invariant operators that will be of interest below.
Meson operators
\eqn\mj{(M_j)_{\tilde i}^i=
\tilde Q_{\tilde i} X^{j-1}Q^i;\;\;\;j=1,2,\cdots, k}
and baryon
operators that are defined as follows.
Introduce ``dressed quarks:''
\eqn\j{Q_{(l)}=X^{l-1}Q;\;\;\;l=1,\cdots k.}
Then construct baryon--like operators
\eqn\baryon{B^{(n_1,n_2,\cdots,n_k)}=Q_{(1)}^{n_1}\cdots Q_{(k)}^{n_k};\;\;
\sum_{l=1}^k n_l=N_c} where the color indices are contracted with an
$\epsilon$ tensor.
The total number of baryon operators of the form \baryon\ is:
\eqn\totnum{\sum_{\{n_l\}}{N_f\choose n_1}\cdots{N_f\choose n_k}=
{kN_f\choose N_c}}

\noindent c) The theory with no superpotential has two independent
$R$ symmetries.
Adding \b\ leaves just one of the two unbroken, with $X$ carrying $R$
charge $2/(k+1)$. When \b\ is relevant, the IR scaling dimensions of
$X$, $Q^i$ are determined by their $R$ charges\foot{Except at very strong
coupling where, as we will see, one can use a dual picture to study the IR
scaling.}.

For $k=1$ \b\ is a mass term for $X$; the adjoint superfield
decouples in the IR and one is led
back to supersymmetric QCD \refs{\ads-\cern}. The case $k=2$
was discussed in \dk.

\newsec{Stability.}

Before turning to duality, we prove,
using an idea that will be useful
later, that the theory described by \b\ has a stable vacuum iff
\eqn\ee{N_f\ge{N_c\over k}.}
Consider a deformation of the superpotential of the theory
\b\ to include lower order terms:
\eqn\bb{W(X)=\Tr\sum_{l=1}^k g_l X^{l+1}+\lambda\Tr X}
We have introduced a Lagrange multiplier $\lambda$ to enforce
the tracelessness condition $\Tr X=0$. Eq.
\bb\ describes a soft perturbation of \b, which does not change the large
field behavior of $W$, hence if there is no stable vacuum for small $g_l$,
the
theory with $g_l=\delta_{l,k} g_k$ has no vacuum either (and vice versa).
Thus, consider the theory with small $g_l$ \bb. The system has multiple
vacua with $Q=\tilde Q=0$ and non vanishing expectation
values of the eigenvalues of $X$. Vacua of the theory are found by
setting the potential for the eigenvalues $x_i$ of $X$ to zero,
$W^\prime(x_i)=0$.
$W^\prime(x)$ is a polynomial of degree $k$,
hence there are $k$
solutions, which are generically distinct.
Ground states are labeled by sequences of integers $i_1\le i_2\le\cdots\le
i_k$,
where $i_l$ is the number of eigenvalues of the matrix $X$
residing in the $l$'th minimum of the potential. Clearly,
\eqn\cc{\sum_{l=1}^ki_l=N_c.}
$\lambda$ is then determined by requiring that the sum
of the eigenvalues (which depend on $\lambda$) vanishes.

In each vacuum $X$ has a quadratic superpotential, i.e. it is massive
and can be integrated out. The gauge group is broken
by the $X$ expectation value:
\eqn\dd{SU(N_c)\to SU(i_1)\times SU(i_2)\times\cdots\times SU(i_k)
\times U(1)^{k-1}}
Some of the $i_l$ may vanish, in which case \dd\ is modified in an
obvious way.
Each of the $SU(i_l)$ factors describes
a supersymmetric QCD model. It is well known \ads,
\cern\ that SQCD has no stable vacuum when the number of flavors is
smaller than the number of colors. Here, this implies that the system has
a stable vacuum iff
\eqn\gg{i_l\le N_f;\;\;\forall\;\;1\le l\le k.}
Eq. \cc\ then implies that a stable vacuum exists iff \ee\ is satisfied.
Finally, taking $g_l\to0$ ($l<k$) in \bb\ we conclude that the same is true
for the theory \b. It would be interesting to analyze the superpotential
that destabilizes the theory when $N_f<N_c/k$ directly, generalizing
the discussion of the case $k=1$ \ads.

By fine tuning the coefficients $g_l$ one can arrange for some of the roots
of $W^\prime$ to coincide. In that case $X$ does not decouple in the
different vacua but rather is governed by a superpotential of the form
\b\ with a lower value of $k$ equal to the order of a particular root of
$W^\prime$. The stability analysis can be repeated for this case too, with
the same conclusions \ee.

\newsec{Duality.}

The anomaly free global symmetry of the $SU(N_c)$ gauge theory
described above is
\eqn\globsym{SU(N_f)\times SU(N_f) \times U(1)_B \times U(1)_R }
with the matter fields transforming as:
\eqn\d{\eqalign{
Q &\qquad (N_f,1,1, 1-{2\over k+1}{N_c\over N_f}) \cr
\tilde Q & \qquad (1, \overline N_f,-1,1-{2\over k+1}{N_c\over N_f})\cr
X &\qquad (1,1,0, {2\over k+1}) .\cr
}}
Following
\nati, \dk\  it is natural to propose a dual theory
with gauge group $SU(kN_f-N_c)$ and the following matter content:
$N_f$ flavors of (dual) quarks $q_i$, $\tilde q^i$, an adjoint field
$Y$, and gauge singlets $M_j$ representing \mj,
$j=1,\cdots, k$, with the transformation properties under the
global symmetry \globsym:
\eqn\e{\eqalign{
q &\qquad (\overline N_f,1,{N_c\over kN_f-N_c}, 1-{2\over
k+1}{kN_f-N_c\over
 N_f})
\cr
\tilde q & \qquad (1, N_f,-{N_c\over kN_f-N_c},1-{2\over k+1}{
kN_f-N_c\over N_f}) \cr
Y &\qquad (1,1,0, {2\over k+1}) \cr
M_j &\qquad (N_f,\overline N_f,0, 2-{4\over k+1}{N_c\over
N_f}+{2\over k+1}(j-1)) \cr
}}
Note again that $j\le k$ since $X^l$ is not an independent chiral operator
for $l\ge k$ \h.
The superpotential in the dual, ``magnetic'', theory is taken to be\foot{
Again, we assume $k< kN_f-N_c$.}:
\eqn\f{W_{\rm mag}=\Tr Y^{k+1}+\sum_{j=1}^k M_j\tilde q Y^{k-j} q.}
For simplicity we set the coeffients in \f\ to one. These coefficients
are calculable and relevant for a more detailed understanding of duality.
One can check using \e\ that $W_{\rm mag}$ preserves the $R$ symmetry
$U(1)_R$.
The case $k=1$ corresponds to the duality of \nati, since $X$, $Y$ are
then massive and can be integrated out. For $k=2$ we recover the case
described in \dk. Below we shall see that theories with different $k$'s
are connected via the flows \bb, so in a sense \e\ generalizes the
previous results.

The `t Hooft anomalies of the dual
theories match, and are given by:
\eqn\g{\eqalign{
SU(N_f)^3 \qquad &N_c d^{(3)}(N_f) \cr
SU(N_f)^2U(1)_R \qquad & -{2\over k+1}{N_c^2 \over N_f}d^{(2)}(N_f) \cr
SU(N_f)^2U(1)_B \qquad & N_cd^{(2)}(N_f) \cr
U(1)_R \qquad &-{2\over k+1}(N_c^2+1) \cr
U(1)_R^3 \qquad &\left(({2\over
k+1}-1)^3+1\right)(N_c^2-1)-{16\over(k+1)^3}
{N_c^4\over N_f^2} \cr
U(1)_B^2U(1)_R \qquad &-{4\over k+1}N_c^2 . \cr}}
The discussion of operator matching proceeds
again along the lines of \nati, \dk:  the mesons
$\tilde Q_{\tilde i} X^{j-1} Q^i$, $j=1,\cdots, k$ \mj\ are as is by now
standard explicitly introduced in the dual theory as additional
gauge singlet fields
$(M_j)^i_{\tilde i}$, \e. $\Tr X^j$, $j=2,\cdots, k$ are mapped to
$\Tr Y^j$.
The mapping of the baryons \baryon\ between the two dual theories
is:
\eqn\barmap{B_{\rm el}^{(n_1,n_2,\cdots,n_k)}\leftrightarrow
B_{\rm mag}^{(m_1,m_2,\cdots,m_k)};\;\;
m_l=N_f-n_{k+1-l};\;\;l=1,2,\cdots, k}
A non-trivial check of duality is the statement
that the charge assignments \d, \e\ necessary for `t Hooft anomaly
matching are also compatible with the map \barmap.

\newsec{Deformations.}

There are many interesting deformations of the theories \b, \f\ that
provide further checks on the duality of the previous section. We will
only discuss two here. The first involves giving a mass to one of the
original, ``electric'', quarks. Thus we add a term to the electric
superpotential \b:
\eqn\mass{W_{\rm el}=g_k\Tr X^{k+1}+m\tilde Q_{N_f} Q^{N_f}}
This gives a mass to $Q^{N_f}$, $\tilde Q_{N_f}$ and reduces the number of
flavors in the IR by one unit keeping $N_c$ fixed: $(N_c, N_f)\to(N_c,
N_f-1)$. Since the theory with $(N_c, N_f-1)$ is dual to one with
$(kN_f-N_c-k, N_f-1)$, we expect that in the dual ``magnetic''
theory \f, \mass\
reduces the number of colors by $k$ units while reducing $N_f$ by one.
The magnetic superpotential is in this case:
\eqn\magmass{W_{\rm mag}=g_k\Tr Y^{k+1}+
\sum_{j=1}^k M_j\tilde q Y^{k-j}q+m(M_1)_{N_f}^{N_f}}
Integrating out the massive fields we find that the vacuum satisfies:
\eqn\const{q_{N_f}Y^{l-1}\tilde q^{N_f}=-\delta_{l,k}m;\;\;l=1,\cdots, k}
which together with some additional conditions fixes the expectation
values:
\eqn\soltn{\eqalign{\tilde q^{N_f}_\alpha=&\delta_{\alpha,1};\cr
                           q_{N_f}^\alpha=&\delta^{\alpha,k};\cr
Y^\alpha_\beta=&\cases{\delta^\alpha_{\beta+1}&$\beta=1,\cdots, k-1$\cr
                        0& otherwise\cr}\cr}}
The Higgs mechanism reduces the number of
colors by $k$ units, and takes $N_f\to N_f-1$. It is not difficult
to extend the discussion to perturbations of the form $m_j\tilde Q_{N_f}
X^{j-1} Q^{N_f}$. These reduce the number of colors in the dual, magnetic,
theory by
$k+1-j$ and give rise (in general)
to a superpotential for the quarks coming from the
reduction of the adjoint field $Y$.

The second deformation we will discuss involves perturbations of the
superpotential \b\ given by \bb. Consider first, for simplicity, the case
$k=2$, where in the electric theory \b:
\eqn\newW{W(X)=\Tr\left(X^3+{m\over2} X^2+\lambda X\right).}
Here $X$ is a general Hermitean matrix,  and
$\lambda$  is
the Lagrange
multiplier introduced in \bb\ to enforce the condition $\Tr X=0$.
Vacuum solutions
are diagonal matrices $X$ with eigenvalues $x_i$ satisfying a quadratic
equation, $3x^2+mx+\lambda=0$. There are two solutions  $x^\pm $
 corresponding to the two minima of the bosonic potential
 $ V=|W^\prime(X)|^2 $.
There are  generically $ N_c+1 $ possible vacua
labeled by $r=0,1,\cdots, N_c$, the number
of eigenvalues  $x_i$ which equal $ x^+ $ the other $ N_c-r $ having the
value $x^-$;   vacua related by
the $Z_2$ operation $r\to N_c-r$ are identical.
The Lagrange
multiplier $\lambda$ is determined by setting $\Tr X=0$,   i.e.
$r x^+(\lambda) + (N_c-r) x^-(\lambda)=0 $ .
The gauge group is broken to:
\eqn\gb{SU(N_c)\to SU(r)\times SU(N_c-r)\times U(1).}
For $r=0, N_c$ $SU(N_c)$ remains unbroken.
As discussed in section 2, in each vacuum the theory reduces to
SQCD ($X$ is massive) and if, without loss of generality, we take
$N_f>N_c$, all $N_c+1$ vacua are stable.

Consider now the dual, magnetic, theory. A similar analysis seems to
suggest
 that
there are $2N_f-N_c+1(>N_c+1)$ vacua with:
\eqn\gbdual{SU(2N_f-N_c)\to SU(l)\times SU(2N_f-N_c-l)\times U(1).}
However, using the results of \ads\ we know that the $l$'th vacuum
is stable iff $l\le N_f$ and $2N_f-N_c-l\le N_f$. Thus,
$l=N_f-N_c,\cdots, N_f$ and there are again $N_c+1$ vacua, as required by
duality. The precise map between the vacua \gb\ and \gbdual\ is
$l=N_f-r$, and the equivalence between the two is the
duality of \nati. In particular, certain linear combinations of
the two gauge singlets $M_1, M_2$ \e\ (which were denoted by $M$, $N$
in \dk) become the meson fields in the two
vacua needed for the duality of \nati.

For $r=0,N_c$ in \gb\ the $SU(N_c)$ gauge group remains unbroken.
It is interesting that the duality described above takes the trivial
electric vacuum $\langle X\rangle=0$ to a magnetic vacuum with $\langle
Y\rangle\not=0$.
In the appropriate vacuum of the magnetic theory, the gauge group is
broken at tree level to:
\eqn\rzero{SU(2N_f-N_c)\to SU(N_f)\times SU(N_f-N_c)\times U(1).}
Denoting the dual quarks of the $SU(N_f)$ sector in \rzero\ by $q, \tilde
q$ and the singlet meson combination that couples to $q, \tilde q$ by
$M$, we have in the $SU(N_f)$ theory the standard \nati\ superpotential
$W=M\tilde q q$. The $M$ equation of motion sets $\tilde q q$ to zero.
But from \natii\ we know that in this theory which has the same number of
colors and flavors, there is a constraint on the quantum moduli space,
$\det\tilde q q-B\tilde B=\Lambda^{2N_f}$, relating the mesons $\tilde q q$
and baryons, $B$, $\tilde B$. Since $\tilde q q=0$, we have $B\tilde
B=-\Lambda^{2N_f}$. The expectation value of $B$ breaks the $U(1)$
symmetry in \rzero\ and again reduces the prediction of the duality of
\dk\ to that of \nati.

For $k>2$ there is a much richer set of deformations of the superpotential
\bb\ connecting the different dualities. Since the analysis is
conceptually similar to the $k=2$ case described above, we only
sketch the structure here.

For generic $W$ of degree $k+1$, there are many vacua found by solving the
 polynomial equation
$W^\prime(X)=0$. As described above, ground states are
labeled by the number of eigenvalues $i_l$ residing in the $l$'th minimum
of the bosonic potential
($l=1,\cdots k$). The
gauge group is broken as in \dd. In the dual theory the situation is
similar
with $j_l$ eigenvalues in the $l$'th minimum, $\sum_l
j_l=kN_f-N_c$, and:
\eqn\dddual{SU(kN_f-N_c)\to SU(j_1)\times SU(j_2)\times\cdots\times SU(j_k)
\times U(1)^{k-1}.}
The different vacua of the two dual theories are mapped to each other by
the duality map,
$j_l=N_f-i_l$.

By fine tuning the coefficients
$g_l$ in \bb\ one can make two or more
roots of $W^\prime$ coincide. This leads in general to:
\eqn\multicr{W^\prime(x)=\prod_i(x-a_i)^{n_i};\;\;\sum n_i=k.}
The theory near $X=a_i$ has $W\sim(x-a_i)^{n_i+1}$. If $r_i$
eigenvalues of $\langle X\rangle$ are equal to $a_i$, the gauge group is
broken to:
\eqn\newbr{SU(N_c)\to\prod_i SU(r_i)\times U(1)^{k-1};\;\;\sum r_i=N_c.}
In the magnetic theory,   in
the corresponding vacua the gauge group is broken as follows:
\eqn\newbrdual{SU(kN_f-N_c)\to\prod_i SU(\tilde r_i)\times
U(1)^{k-1};\;\;\sum \tilde r_i=kN_f-N_c.}
It is easy to show using the fact that stable vacua of a theory with
superpotential $W=X^{k+1}$ exist only for $N_f\ge N_c/k$ (see section 2),
that there is a
one to one correspondence of the vacua \newbr, \newbrdual\ with
$\tilde r_i=n_iN_f-r_i$.
We see that the duality transformation \d, \e\ with a certain $k$ gives
rise after perturbing the superpotential as in \bb\ to products of
theories dual under the same duality with smaller values of $k$. The
perturbations \bb\ therefore connect the different dualities. The
consistency of the resulting picture is further evidence for the duality
of the previous section.

\newsec{Comments.}

1) At the self dual points of the duality transformations discussed
above, $N_f=2N_c/k$ many new operators  with R charge two
appear.
Examples include $M_j M_{k+1-j}$, $j=0,\cdots,[{k+1\over2}]$
($M_j$ are defined in \mj; $M_0\equiv 1$).
It is likely
\ref\ls{R. Leigh and M. Strassler, RU-95-2, hep-th/9503121}\
that these operators are actually exactly marginal  in the IR conformal
field theory
and lead to manifolds of fixed points. The appearance of new marginal
operators at self dual points seems to be a very general phenomenon in
four dimensional duality, and is reminiscent of similar phenomena in two
dimensional theories, where at self dual points one usually encounters
enhanced symmetries and new moduli.

The duality described in the previous sections should act in this case
on the manifold of fixed points described by the superpotential:
\eqn\wfll{W=g_k\Tr X^{k+1}+{1\over2}\sum_{j=1}^k
 \lambda_j M_j M_{k+1-j}}
$\lambda_j(=\lambda_{k+1-j})$ are coordinates
on the moduli space of IR fixed points.
Duality presumably interchanges large and small $\lambda_j$,
and electric and magnetic variables. One can generalize
the discussion of \ls\ to study some aspects of this duality, such as the
appearance of the singlet mesons in the magnetic theory \e. Rewrite
\wfll\ as ($h_j=h_{k+1-j}$):
\eqn\wmod{W=g_k\Tr X^{k+1}-
\sum_{j=1}^k
\left(
{h_j^2\over2\lambda_j} N_j N_{k+1-j}
+h_jN_jM_{k+1-j}\right)}
in terms of auxiliary fields $(N_j)_i^{\tilde i}$, which become
dynamical at large distances. In the limit $\lambda_j\to0$
we see from
\wfll, \wmod\ that the theory approaches the electric theory
described above \b.
As $\lambda_j\to\infty$ the mass of $N_j$ goes to zero, and the theory
approaches the magnetic theory \e, \f, with $N_j$ playing the role
of the gauge singlet mesons in \e. For generic $\lambda_j$
the singlet mesons $N_j$ are massive,     the global symmetry \globsym\
$SU(N_f)\times SU(N_f)$ is broken to $SU(N_f)$, and the anomaly matching
does not require elementary ``meson'' chiral superfields .
The full symmetry \globsym\ is restored at $\lambda_j=0,\infty$.

One can also study the exactly marginal deformation
 induced by the operator corresponding to $j=0$ in \wfll:
          \eqn\wfl{W_{fl}=g_k\Tr X^{k+1}+h_k\tilde Q_i X^kQ^i.}
For $k=1$, $g_k=0$ \wfl\ describes the line
of fixed points of finite $\cn=2$ supersymmetric
theories with $N_f=2N_c$ \sw. For $k>1$, $g_k=0$ one can think of \wfl\
as describing the finite $\cn=2$ model with all but $2N_c/k$ flavors given
masses and integrated out \ls.
Thus, it is possible that the structure described in this paper is
related to   $\cn=2$ duality \sw.

2) Our results suggest the following picture for the theory with $W=0$.
As $N_f$ decreases from $2N_c$ the IR dimension of $X$ becomes smaller,
so that $X^{k+1}$ become relevant for any $k <N_c$, when
$N_f<N_0(k)$. In this regime, adding the superpotential \b\ takes the
theory to the fixed point we have discussed here, which is distinct from
the $W=0$ one. Duality suggests that $N_0(k)>2N_c/k$, so that at
the self dual point $N_f=2N_c/k$, $\Tr X^{k+1}$ is strongly
relevant. For $N_f>N_0(k)$ the $W=0$ fixed point is unique,
ignoring deformations and restrictions of the chiral ring.
Duality implies a mirror image of this picture for
$N_f<2N_c/k$. Calculating $N_0(k)$ is tantamount to
calculating the scaling dimension of $X$ in the $W=0$ theory and would
be an important clue to the structure of the theory.

3) For large $k$ the scaling dimension of $X$ at the fixed point
governed by \b\ is small and the dimensions of certain gauge
invariant operators like $\Tr X^2$ are not governed by their $R$ charges,
due to unitarity bounds
\ref\mack{G. Mack, \cmp{55}{1977}{1}}.
Consider the theory
with no superpotential, $W=0$. As discussed above, when $N_f$ decreases,
the dimension of $X$ decreases. At a certain $N_f=N_2$, the dimension of
$\Tr X^2$ descends to one. The operator becomes a free field  \mack, and
decouples from the dynamics. This remains the case for $N_f<N_2$ and
clearly also when we turn on the superpotential \b. At some lower value,
$N_f=N_3$ the same happens to $\Tr X^3$, etc. By duality, these operators
``recouple'' at some lower values of $N_f<2N_c/k$, which are more
conveniently studied in the dual, magnetic theory. The role of the
decoupled, free fields deserves further investigation. They should
play an important role in a dual description of the theory with no
superpotential, perhaps appearing as elementary
fields in such a description.
\bigskip
\noindent {\bf Note added:} Some related issues are
discussed in a recent preprint
\ref\ta{O. Aharony, J. Sonnenschein and S. Yankielowicz, TAUP-2246-95,
hep-th/9504113}.

\bigskip

\bigskip

\centerline{{\bf Acknowledgments}}

We thank N. Seiberg for collaboration on some of the results, and
E. Martinec for discussions.
The work  of D.K.
was supported in part by a DOE OJI award, that of A.S. by BSF grant
number 5360/2 and by the Minerva foundation.

\listrefs
\bye